\documentclass{llncs}
\usepackage{comment}
\usepackage{url}
\usepackage{amssymb} 
\usepackage{amsmath} 
\usepackage{hhline}
\usepackage{paralist}
\usepackage{stmaryrd} 
\usepackage{url}
\usepackage{multirow}
\usepackage{times}
\usepackage{helvet}
\usepackage{courier}
\usepackage{graphicx}
\usepackage{algorithm}
\usepackage{algorithmic}
\usepackage{hyperref}
\usepackage{caption}

\newcommand{\var}[1]{\mathit{vars}(#1)}
\newcommand{\LL}{\mathcal{L}}

\newcommand{\G}{\mathcal{G}}

\newcommand{\andd}{\mathbin{\mathrm{AND}}}
\newcommand{\OPT}{\mathbin{\mathrm{OPT}}}

\newcommand{\UNION}{\mathbin{\mathrm{UNION}}}

\newcommand{\FILTER}{\mathbin{\mathrm{FILTER}}}
\newcommand{\SELECT}{\mathop{\mathrm{SELECT}}}

\newcommand{\Trace}{\mathit{Trace}}

\newcommand{\voc}{\mathit{adom}}

\newcommand{\axis}{\mathit{axis}}
\newcommand{\self}{\mathit{self}}

\newcommand{\nexts}{\mathit{next}}

\newcommand{\sem}[1]{\llbracket #1 \rrbracket_{G}}
\newcommand{\semm}[2]{\llbracket #1 \rrbracket_{#2}}

\newcommand{\edges}{\mathit{edge}}
\newcommand{\nodes}{\mathit{node}}
\newcommand{\dom}[1]{\mathrm{dom}(#1)}

\newcommand{\vocvertor}{\mathit{Convertor}}

\usepackage{tikz}
\usetikzlibrary{arrows,positioning,automata,decorations,fit,backgrounds,calc,shapes,snakes}
\usepgflibrary{shapes.geometric} 
\usetikzlibrary{shapes.geometric} 
\usepgflibrary{decorations.pathmorphing} 
\usetikzlibrary{decorations.pathmorphing} 
\usepackage{verbatim}
\usetikzlibrary{calc,backgrounds}
\usetikzlibrary{trees}

\begin{document}

\title{Context-Free Path Queries on RDF Graphs}
\author{Xiaowang Zhang \and Zhiyong Feng \and Xin Wang \and Guozheng Rao \and Wenrui Wu}
\institute{School of Computer Science and Technology, Tianjin University, China\\
Tianjin Key Laboratory of Cognitive Computing and Application, Tianjin,
China\\ $\{$xiaowangzhang, zyfeng, wangx, rgz, wenruiwu$\}${@}tju.edu.cn}

\maketitle
\begin{abstract}
  Navigational graph queries are an important class of queries that
  can extract implicit binary relations over the nodes of input
  graphs. Most of the navigational query languages used in the RDF
  community, e.g. property paths in W3C SPARQL 1.1 and nested regular
  expressions in nSPARQL, are based on the regular expressions. It is
  known that regular expressions have limited expressivity; for
  instance, some natural queries, like \textit{same generation-queries}, are not expressible with regular
  expressions. To overcome this limitation, in this paper, we present
  {cfSPARQL}, an extension of SPARQL query language equipped with
  context-free grammars. The {cfSPARQL} language is strictly more
  expressive than property paths and nested expressions. The
  additional expressivity can be used for modelling graph
  similarities, graph summarization and ontology alignment.
  Despite the increasing expressivity, we show that cfSPARQL still
  enjoys a low computational complexity and can be evaluated
  efficiently.
\end{abstract}

\section{Introduction}\label{sec:intro}
The Resource Description Framework (RDF) \cite{rdfprimer} recommended by
World Wide Web Consortium (W3C) is a standard graph-oriented model for data
interchange on the Web \cite{gutierrez_survey}. RDF has a broad range of
applications in the semantic web, social network, bio-informatics,
geographical data, etc \cite{abs_book}. Typical access to
graph-structured data is its navigational nature
\cite{DBLP:conf/icdt/HellingsKBZ13,DBLP:conf/pods/LibkinRV13,DBLP:journals/isci/FletcherGLSBGVW15}.
Navigational queries on graph databases return binary relations over the
nodes of the graph \cite{barcelo_crpq_pods}. Many existing navigational
query languages for graphs are based on binary relational algebra such as
XPath (a standard navigational query language for trees
\cite{marxrijke_xpath}) or regular expressions such as RPQ (regular path
queries) \cite{DBLP:conf/icdt/Reutter0V15}.

SPARQL \cite{sparql} recommended by W3C has become the standard language for
querying RDF data since 2008 by inheriting classical relational languages
such as SQL. However, SPARQL only provides limited navigational
functionalities for RDF \cite{nsparql,Taski_SPARQL}. Recently, there are
several proposed languages with navigational capabilities for querying RDF
graphs
\cite{versa,DBLP:conf/esws/KochutJ07,DBLP:conf/www/AnyanwuMS07,nsparql,DBLP:journals/ws/AlkhateebBE09,DBLP:conf/swws/AlkhateebBE08,DBLP:journals/ijwis/AlkhateebE14,DBLP:conf/aaai/FiondaPC15,DBLP:conf/semweb/KostylevR0V15}.
Roughly, Versa \cite{versa} is the first language for RDF with navigational
capabilities by using XPath over the XML serialization of RDF graphs.
SPARQLeR proposed by Kochut et al. \cite{DBLP:conf/esws/KochutJ07}  extends
SPARQL by allowing path variables. SPARQL2L proposed by Anyanwu et al.
\cite{DBLP:conf/www/AnyanwuMS07} allows path variables in graph patterns and
offers good features in nodes and edges such as constraints.
PSPARQL proposed by Alkhateeb et al. \cite{DBLP:journals/ws/AlkhateebBE09}
extends SPARQL by allowing regular expressions in general triple patterns
with possibly blank nodes and CASPAR further proposed by Alkhateeb et al.
\cite{DBLP:conf/swws/AlkhateebBE08,DBLP:journals/ijwis/AlkhateebE14} allows
constraints over regular expressions in PSPARQL where variables are allowed
in regular expressions. nSPARQL proposed by P\'erez et al. \cite{nsparql}
extends SPARQL by allowing nested regular expressions in triple patterns.
Indeed, nSPARQL is still expressible in SPARQL if the transitive closure relation is
absent \cite{Taski_SPARQL}. In March 2013, SPARQL 1.1 \cite{sparql1.1}
recommended by W3C allows property paths which strengthen the navigational
capabilities of SPARQL1.0
\cite{DBLP:conf/aaai/FiondaPC15,DBLP:conf/semweb/KostylevR0V15}.

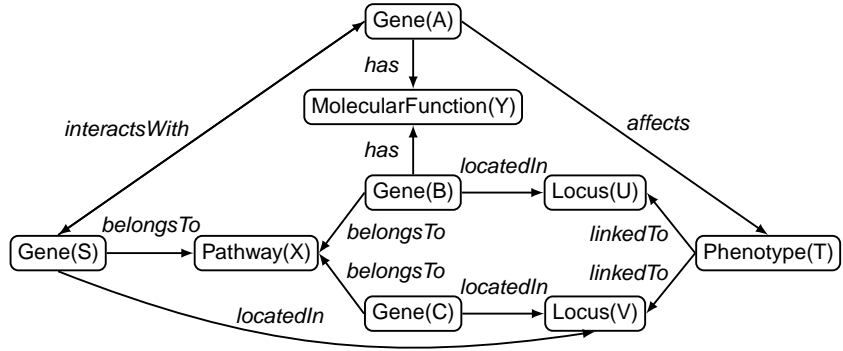
\begin{figure}[t]
\setlength{\belowcaptionskip}{-0.5cm}
\begin{center}
\scalebox{0.9}{
\begin{tikzpicture}[>=latex]

  %
  %
  \tikzstyle{state} = [draw, thick, fill=white, rectangle, rounded corners=.8ex, minimum height=1.5em, minimum width=1.5em, node distance=4em, font={\sffamily}]
  \tikzstyle{stateEdgePortion} = [black,thick];
  \tikzstyle{stateEdge} = [stateEdgePortion, ->];
  \tikzstyle{stateEdgeD} = [stateEdgePortion,dashed,->];
  \tikzstyle{stateEdgeUD} = [stateEdgePortion,dashed, <->];
  \tikzstyle{edgeLabel} = [pos=0.5, text centered, font={\sffamily\small}];

  %
  %
  \node[state, name=p1] {Gene(A)};
  \node[state, name=p2, below of=p1] {MolecularFunction(Y)};
  \node[state, name=p3, below of=p2] {Gene(B)};
  \node[state, name=p5, below left of=p3,  xshift=-4.5em] {Pathway(X)};
  \node[state, name=p6, left of=p5, xshift=-5.25em] {Gene(S)};
  \node[state, name=p4, below right of=p5,  xshift=4.5em] {Gene(C)};
  \node[state, name=p7, right of=p3,  xshift=4.5em] {Locus(U)};
  \node[state, name=p9, below right of=p7,  xshift=5.25em] {Phenotype(T)};
  \node[state, name=p8, below left of=p9,  xshift=-5.25em] {Locus(V)};

 \draw ($(p1.south)$) edge[stateEdge] node[edgeLabel, xshift=-1.5em]{\emph{has}} ($(p2.north)$);
 \draw ($(p1.east)$) edge[stateEdge] node[edgeLabel, xshift=2em, yshift=0.25em]{\emph{affects}} ($(p9.north)$);
 \draw ($(p3.north)$) edge[stateEdge] node[edgeLabel, xshift=-1.5em]{\emph{has}} ($(p2.south)$);
 \draw ($(p3.west)$) edge[stateEdge] node[edgeLabel, xshift=2.5em, yshift=-0.5em]{\emph{belongsTo}} ($(p5.east)$);
 \draw ($(p6.east)$) edge[stateEdge] node[edgeLabel, yshift=1.25em]{\emph{belongsTo}} ($(p5.west)$);
 \draw ($(p4.west)$) edge[stateEdge] node[edgeLabel, xshift=2.5em, yshift=0.5em]{\emph{belongsTo}} ($(p5.east)$);
 \draw ($(p3.east)$) edge[stateEdge] node[edgeLabel, yshift=1.25em]{\emph{locatedIn}} ($(p7.west)$);
 \draw ($(p9.west)$) edge[stateEdge] node[edgeLabel, xshift=-2em, yshift=-0.5em]{\emph{linkedTo}} ($(p7.east)$);
 \draw ($(p9.west)$) edge[stateEdge] node[edgeLabel, xshift=-2em, yshift=0.5em]{\emph{linkedTo}} ($(p8.east)$);
 \draw ($(p4.east)$) edge[stateEdge] node[edgeLabel, yshift=1.25em]{\emph{locatedIn}} ($(p8.west)$);
 \draw ($(p6.south)$) edge[stateEdge, bend right=15] node[edgeLabel, xshift=-2em, yshift=1.25em]{\emph{locatedIn}} ($(p8.south)$);
 \draw ($(p6.north)$) edge[stateEdge] node[edgeLabel, xshift=-4em]{\emph{interactsWith}} ($(p1.west)$);
 \draw ($(p1.west)$) edge[stateEdge] ($(p6.north)$);
\end{tikzpicture}
}
\caption{A biomedical ontology \cite{sevonCFG}}\label{fig:toy}
\end{center}
\vspace{-2mm}
\end{figure}

However, those regular expression-based extensions of SPARQL are still
limited in representing some more expressive navigational queries which are
not expressed in regular expressions. Let us consider a fictional biomedical
ontology mentioned in \cite{sevonCFG} (see Figure \ref{fig:toy}). We are
interested in a navigational query about those paths that confer similarity
(e.g., between \emph{Gene(B)} and \emph{Gene(C)}), which suggests a causal
relationship (e.g., between \emph{Gene(S)} and \emph{Phenotype(T)}). This
query about similarity arises from the well-known \emph{same
generation-query} \cite{ahv_book}, which is proven to be inexpressible in any regular
expression. To express the query, we have to introduce a query embedded with a context-free grammar (CFG) for expressing the language of $\{ww^{T} \mid w \emph{ is a string}\}$\cite{sevonCFG} where $w^{T}$ is the converse of $w$. For instance, if $w$ = ``$abcdfe$'' then $w^{T}$ = ``$e^{-1}f^{-1}d^{-1}c^{-1}b^{-1}a^{-1}$''. As we know, CFG has more expressive power than any regular expression \cite{hu}. Moreover, the context-free grammars can provide a simplified more user-friendly dialect of Datalog \cite{abs_book} which still allows powerful recursion\cite{hu}. Besides, the context-free graph queries have also practical query evaluation strategies.
For instance, there are some applications in verification \cite{LangeMC}. So it is
interesting to introduce a navigational query embedded with context-free grammars
to express more practical queries like the same generation-query.

A proposal of conjunctive context-free path queries (written by
\emph{Helling's CCFPQ}) for edge-labeled directed graphs has been presented by
Helling \cite{DBLP:conf/icdt/Hellings14} by allowing context-free grammars in
path queries. A naive idea to express same generation-queries is
transforming this RDF graph to an edge-labeled directed graph via navigation
axes \cite{nsparql} and then using Helling's CCFPQ since an RDF graph can be
intuitively taken as an edge-labeled directed graph. However, this
transformation is difficult to capture the full information of this RDF graph
since there exist some slight differences between RDF graphs and
edge-labeled directed graphs, particularly regarding the connectivity
\cite{DBLP:conf/semweb/HayesG04}, thus it could not express some regular
expression-based path queries on RDF graphs. For instance, a nested regular
expression (nre) of the form $\axis::[e]$ on RDF graphs in nSPARQL
\cite{nsparql}, is always evaluated to the empty set over any
edge-labeled directed graph. That is to say, an nre of the form ``$\axis::[e]$'' is
hardly expressible in Helling's CCFPQ.

To represent more expressive queries with efficient query evaluation is a renewed interest topic in the classical topic of graph databases \cite{ahv_book}.
Hence, in this paper, we present a context-free extension of path queries and SPARQL queries on RDF graphs which can express both nre and nSPARQL \cite{nsparql}. Furthermore, we study several fundamental properties of the proposed context-free path queries and context-free SPARQL queries. The main contributions of this paper can be summarized as follows:
\begin{compactitem}
\item We present \emph{context-free path queries} (CFPQ) (including \emph{conjunctive context-free path queries}  (CCFPQ),  \emph{union of simple conjunctive context-free path queries} \\ (UCCFPQ$^{s}$), and \emph{union of conjunctive context-free path queries} (UCCFPQ) for RDF graphs and find that CFPQ, CCFPQ, and UCCFPQ have efficient query evaluation where the query evaluation has the polynomial data complexity and the NP-complete combined complexity. Finally, we implement our CFPQs and evaluate experiments on some popular ontologies.
\item We discuss the expressiveness of CFPQs by referring to nested regular expressions (nre). We show that CFPQ, CCFPQ, UCCFPQ$^{s}$, and UCCFPQ exactly express four fragments of nre, basic nre ``nre$_{0}$'', union-free nre ``nre$_{0}(\mathrm{N})$'', nesting-free nre ``nre$_{0}(|)$'', and full nre, respectively (see Figure \ref{fig:overviewofCFPQ}). The query evaluation of cfSPARQL has the same complexity as SPARQL.
\item We propose \emph{context-free SPARQL} (cfSPARQL) and \emph{union of
    conjunctive context-free SPARQL} (uccfSPARQL) based on CFPQ and UCCFPQ,
    respectively. It shows that cfSPARQL has the same expressiveness as that of
    uccfSPARQL. Furthermore, we prove that cfSPARQL can strictly express
    both SPARQL and nSPARQL (even nSPARQL$^\neg$: a variant of nSPARQL by
    allowing nre with negation ``nre$^{\neg}$) (see Figure
    \ref{fig:overviewofcfSPARQL}).
\end{compactitem}

\paragraph{Organization of the paper}
Section~\ref{sec:preliminaries} recalls nSPARQL and context-free grammar.
Section~\ref{sec:CFPQ} defines CFPQ. Section~\ref{sec:expressivity}
discusses the expressiveness of CFPQ. Section~\ref{sec:cfSPARQL} presents
cfSPARQL and Section \ref{sec:nre_neg} discusses the relations on nre with
negation. Section \ref{sec:implement} \label{sec:implement} evaluates experiments. We conclude
in Section~\ref{sec:conclusion}. Due to the space limitation, all proofs and
some further preliminaries are omitted but they are available in an extended
technical report in arXiv.org \cite{cfSPARQL}.

\section{Preliminaries}\label{sec:preliminaries}
In this section, we introduce the language nSPARQL and context-free grammar.
\subsection{The syntax and semantics of nSPARQL}
In this subsection, we recall the syntax and semantics of nSPARQL, largely
following the excellent expositions \cite{nsparql,perez_sparql_tods}.

\paragraph{\textbf{RDF graphs}}
An RDF statement is a \emph{subject-predicate-object} structure, called
\emph{RDF triple} which represents resources and the properties of those
resources. For the sake of simplicity similar to \cite{nsparql}, we assume
that RDF data is composed only IRIs\footnote{A standard RDF data is composed
of IRIs, blank nodes, and literals. For the purposes of this paper, the
distinction between IRIs and literals will not be important.}. Formally, let
${U}$ be an infinite set of \emph{IRIs}. A triple $(s, p, o) \in {U} \times
{U} \times {U}$ is called an \emph{RDF triple}.  An \emph{RDF graph} $G$ is
a finite set of RDF triples. We use $\voc(G)$ to denote the \emph{active
domain} of $G$, i.e., the set of all elements from ${U}$ occurring in $G$.

For instance, a biomedical ontology shown in Figure \ref{fig:toy} can be modeled in an RDF graph named as $G_{\mathrm{bio}}$ where each labeled-edge of the form $a \overset{p}{\to} b$ is directly translated into a triple $(a, p, b)$.

\paragraph{\textbf{Paths and traces}}
Let $G$ be an RDF graph. A \emph{path} $\pi = (c_1c_2 \ldots c_m)$ in $G$ is
a non-empty finite sequence of constants from $G$, where, for every $i \in
\{1, \ldots, m-1\}$, $c_i$ and $c_{i+1}$ exactly occur in the same triple of $G$
(i.e., $(c_{i}, c, c_{i+1})$, $(c_{i}, c_{i+1}, c)$, and $(c, c_{i},
c_{i+1})$ etc.). Note that the precedence between $c_i$ and $c_{i+1}$ in a
path is independent of the positions of $c_i, c_{i+1}$ in a triple.

In nSPARQL, three different \emph{navigation axes}, namely, $\nexts$,
$\edges$, and $\nodes$, and their inverses, i.e., $\nexts^{-1}$,
$\edges^{-1}$, and $\nodes^{-1}$, are introduced to move through an RDF
triple $(s, p, o)$ \cite{nsparql}.

Let $\Sigma=\{\axis, {\axis::c} \mid c \in {U}\}$ where $\axis \in \{\self$,
$\nexts$, $\edges$, $\nodes$, $\nexts^{-1}$, $\edges^{-1}$,
$\nodes^{-1}\}$. Let $G$ be an RDF graph. We use $\Sigma(G)$ to denote the
set of all symbols $\{\axis, {\axis::c} \mid c \in \voc(G)\}$ occurring in
$G$.

Let $\pi = (c_1\ldots c_m)$ be a path in $G$. A
\emph{trace} of path $\pi$ is a string over $\Sigma(G)$ written by
$\mathcal{T}(\pi) = l_1\ldots l_{m-1}$ where, for all $ i \in \{1, \ldots,
m-1\}$, $(c_ic_{i+1})$ is labeled by $l_i$ and $l_i$ is of the form $\axis$,
$\axis::c$, $\axis^{-1}$, or $\axis^{-1}::c$ \cite{nsparql}. We use $\Trace(\pi)$ to denote the set of all traces of $\pi$.

Note that it is possible that a path has multiple traces since any two nodes possibly
occur in the multiple triples. For example, consider an {RDF} graph $G = \{(a, b,
c), (a, c, b)\}$ and given a path $\pi = (abc)$, both
$(\edges::c)(\nodes::a)$ and $(\nexts::c)(\nodes^{-1}::a)$ are traces of
$\pi$.

For instance, in the RDF graph $G_{\mathrm{bio}}$ (see Figure \ref{fig:toy}), a path from \textit{Gene(B)} to \textit{Gene(C)} has a trace: $(\nexts::\emph{locatedIn})(\nexts^{-1}::\emph{linkedTo})(\nexts::\emph{linkedTo})(\nexts^{-1}::\emph{locatedIn})$.

\paragraph{\textbf{Nested regular expressions}}
Nested regular expressions (\emph{nre}) are defined by the following formal
syntax:
\begin{equation*}\label{def:nre}
e:= \axis \mid \axis::c\, (c \in U) \mid \axis::[e] \mid e /e \mid e | e  \mid  e^{\ast}.
\end{equation*}
Here the \emph{nesting nre} is of the form $\axis::[e]$.

For simplification, we denote some interesting fragments of nre as follows:
\begin{compactitem}
\item nre$_{0}$: \emph{basic nre}, i.e., nre only consisting of ``$\axis$'',
    ``$/$'', and ``$\ast$'';
\item nre$_{0}(|)$: basic nre by adding the operator ``$\mid$'';
\item nre$_{0}(\mathrm{N})$ to basic nre by adding nesting nre
    $\axis::[e]$.
\end{compactitem}


\paragraph{\textbf{Patterns}}
Assume an infinite set ${V}$ of \emph{variables}, disjoint from
${U}$. A \emph{nested regular expression triple} (or \emph{nre-triple}) is a tuple of the form $(?x, e, ?y)$, where $?x, ?y \in V$ and $e$ is an nre\footnote{In nSPARQL
\cite{nsparql}, nre-triples allow a general form $(v, e, u)$ where $u, v
\in U \cup V$. In this paper, we mainly consider the case $u, v\in V$ to
simplify our discussion.}.

Formally, nSPARQL (graph) patterns are recursively constructed from
nre-triples:
\begin{compactitem}
\item An nre-triple is an nSPARQL pattern;
\item All $P_1\, \UNION\, P_2$, $P_1\, \andd\, P_2$, and $P_1\, \OPT\, P_2$
    are nSPARQL patterns if $P_1$ and $P_2$ are nSPARQL patterns;
\item $P\, \FILTER\, C$ if $P$ is an nSPARQL pattern and $C$ is a
    constraint;
\item $\SELECT_{S}(P)$ if $P$ is an nSPARQL pattern and $S$ is a set of
    variables.
\end{compactitem}

%

\paragraph{\textbf{Semantics}}

Given an RDF graph $G$ and an nre $e$, the evaluation of $e$ on $G$, denoted by $\semm
{e}{G}$, is a binary relation. More details can be found in \cite{nsparql}.
Here, we recall the semantics of nesting nre of the form $\axis::[e]$ as
follows: $$\semm {\axis::[e]}{G} = \{(a, b) \mid \exists\, c, d \in
\voc(G),\, (a, b) \in \sem{\axis::c}  \text{ and }(c, d) \in \sem {e}\}.$$

The semantics of nSPARQL patterns is defined in terms of sets of so-called
\emph{mappings}, which are simply total functions $\mu \colon S \to U$ on
some finite set $S$ of variables. We denote the domain $S$ of $\mu$ by $\dom
\mu$.

Basically, the semantics of an nre-triple $(u, e, v)$ is defined as follows:
\begin{equation*}
\sem {(u,e,v)} = \{\mu \colon \{u,v\}\cap V \to U \mid (\mu(u),\mu(v)) \in \semm{e}{G}\}.
\end{equation*}
Here, for any mapping $\mu$ and any constant $c \in {U}$, we agree that
$\mu(c)$ equals $c$ itself.

Let $P$ be an nSPARQL pattern, the semantics of $P$ on $G$, denoted by $\semm P G$,
is analogously defined as usual following the semantics of SPARQL
\cite{nsparql,perez_sparql_tods}.

\paragraph{Query evaluation}

A \emph{SPARQL (SELECT) query} is an nSPARQL pattern. Given a RDF graph $G$, a
pattern $P$, and a mapping $\mu$, the query evaluation problem of nSPARQL is to decide whether $\mu$ is in $\semm {P}{G}$. The complexity of query evaluation problem
is PSpace-complete \cite{perez_sparql_tods}.

\subsection{Context-free grammar}
In this subsection, we recall context-free grammar. For more details, we refer the interested readers to some references about formal languages \cite{hu}.

A \emph{context-free grammar} (COG) is a 3-tuple $\G = (N, A, R)$ \footnote{We deviate from the usual definition of context-free grammar by not including a special start non-terminal following \cite{DBLP:conf/icdt/Hellings14}.} where
\begin{compactitem}
\item $N$ is a finite set of variables (called \emph{non-terminals});
\item $A$ is a finite set of constants (called \emph{terminals});
\item $R$ is a finite set of production rules $r$ of the form $v \to S$, where $v \in N$ and $S \in (N \cup A)^{\ast}$ (the asterisk $\ast$ represents the Kleene star operation). We write $v \to \epsilon$ if $\epsilon$ is the empty string.
\end{compactitem}

A string over $N \cup A$ can be written to a new string over $N \cup A$ by applying production rules. Consider a string $avb$ and a production rule $r: v \to avb$, we can obtain a new string $aavbb$ by applying this rule $r$ one time and another new string $aaavbbb$ by applying the rule $r$ twice. Analogously, strings with increasing length can be obtained in this rule.

Let $S,T \in (N \cup A)^{\ast}$. We write $(S \stackrel{\G}{\rightarrow} T)$ if $T$ can be obtained from $S$ by applying production rules of $\G$ within a finite number of times.

The \emph{language} of grammar $\G = (N, A, R)$ w.r.t. start non-terminal $v \in N$ is defined by $\LL(\G_{v}) = \{S$ a finite string over $A$ $\mid  v \stackrel{\G}{\rightarrow} S\}$.

For example, $\G = (N, A, R)$ where $N = \{v\}$, $A = \{a, b\}$, and $R = \{v \to ab, v \to avb\}$. Thus $\LL(\G_{v}) = \{a^nb^n \mid n \ge 1\}$.

\section{Context-free path queries}\label{sec:CFPQ}
In this section, we introduce context-free path queries on RDF graphs based on context-free path queries on directed graphs \cite{DBLP:conf/icdt/Hellings14} and nested regular expressions \cite{nsparql}.

\subsection{Context-free path queries and their extensions}
In this subsection, we firstly define \emph{conjunctive context-free path queries} on RDF graphs and then present some variants (it also can been seen as extensions).

\paragraph{\bf Conjunctive context-free path queries}

In this paper, we assume that $N \cap V = \emptyset$ and $A \subseteq \Sigma$ for all CFG $\G = (N, A, R)$.

\begin{definition}\label{def:ccfpq}
Let $\G = (N, A, R)$ be a CFG and $m$ a positive integer. A \emph{conjunctive context-free path query} (\emph{CCFPQ})  is of the form $\mathbf{q}(?x,?y)$\footnote{In this paper, we simply write a conjunctive query as a Datalog rule.}, where,
\begin{equation}\label{equ:CCFPQ}
\mathbf{q}(?x, ?y):= \bigwedge^{m}_{i=1}\, \alpha_i,
\end{equation}
where
\begin{compactitem}
\item $\alpha_i$ is a triple pattern either of the form $(?x, ?y, ?z)$ or of the form $v(?x, ?y)$;
\item $\{?x, ?y\} \subseteq \var{\mathbf{q}}$ where $\var{\mathbf{q}}$ denotes a collection of all variables occurring in the body of $\mathbf{q}$;
\item $\{v_1, \ldots, v_m\} \subseteq N$.
\end{compactitem}
We regard the name of query $\mathbf{q}(?x, ?y)$ as $\mathbf{q}$ and call the right of Equation (\ref{equ:CCFPQ}) as the \emph{body} of $\mathbf{q}$.
\end{definition}

\begin{remark}
In our CCFPQ, we allow a triple pattern of the form $(?x, ?y, ?z)$ to characterize those queries w.r.t. ternary relationships such as nre-triple patterns of nSPARQL \cite{nsparql} to be discussed in Section \ref{sec:expressivity}. The formula $v(?x, ?y)$ is used to capture context-free path queries \cite{DBLP:conf/icdt/Hellings14}.
\end{remark}

We say a \emph{simple conjunctive context-free path query} (written by $CCFPQ^{s}$) if only the form $v(?x, ?y)$ is allowed in the body of a CCFPQ. We also say a \emph{context-free path query} (written by \emph{CFPQ}) if $m=1$ in the body of a CCFPQ$^{s}$.

Semantically, let $\G = (N, A, R)$ be a CFG and $G$ an RDF graph, given a CCFPQ $\mathbf{q}(?x, ?y)$ of the form (\ref{equ:CCFPQ}), $\semm {\mathbf{q}(?x, ?y)} {G}$ is defined as follows:
\begin{multline}\label{equ:CCFPQ-semantics}
\{\mu|_{\{?x, ?y\}} \mid \dom{\mu} = \var{\mathbf{q}} \text{ and } \forall\, i =1, \ldots, m,  \mu|_{\var{\alpha_i}} \in \semm {\alpha_i}{G}\},
\end{multline}
where the semantics of $v(?x, ?y)$ over $G$ is defined as follows:
\begin{multline*}
\semm {v(?x, ?y)}{G} = \{\mu \mid \dom{\mu} = \{?x, ?y\} \text{ and }\\ \exists\, \pi= (\mu(?x) c_1 \ldots c_m \mu(?y)) \text{ a path in } G, \Trace(\pi)\cap \LL(\G_{v}) \neq \emptyset\}.
\end{multline*}

Intuitively, $\semm {v(?x, ?y)}{G}$ returns all pairs connected by a path in $G$ which contains a trace belonging to the language generated from this CFG starting at non-terminal $v$.

\begin{example}\label{exam:bio}
Let $\G = (N, A, R)$ be a CFG where $N = \{u, v\}$, $A = \{\nexts::\emph{locatedIn}$, $ \nexts^{-1}::\emph{locatedIn}$, $\nexts::\emph{linkedTo}, \nexts^{-1}::\emph{linkedTo}\}$, and $P = \{v \to (\nexts::\emph{locatedIn})\, u\,(\nexts^{-1}::\emph{locatedIn})$, $u \to (\nexts^{-1}::\emph{linkedTo})\, u\,(\nexts::\emph{linkedTo}), u \to \epsilon\}$. Consider a CFPQ $\mathbf{q}$ be of the form $v(?x, ?y)$. The query $\mathbf{q}$ represents the relationship of similarity (between two genes) since $\LL(\G_{v}) = \{{(\nexts^{-1}::\emph{locatedIn})}^n(\nexts^{-1}::\emph{linkedTo})(\nexts::\emph{linkedTo}){(\nexts::\emph{locatedIn})}^n \mid n\ge 1\}$. Consider the RDF graph $G_{\mathrm{bio}}$ in Figure \ref{fig:toy}, $\semm {\mathbf{q}(?x, ?y)} {G_{\mathrm{bio}}} = \{(?x =\emph{Gene(B)}, ?y = \emph{Gene(C)})\}$. Clearly, the query $\mathbf{q}$ returns all pairs with similarity.
\end{example}

\paragraph{\bf Query evaluation}

Let $\G = (N, A, R)$ be a CFG and $G$ an RDF graph. Given a CCFPQ $\mathbf{q}(?x, ?y)$ and a tuple $\mu = (?x = a, ?y = b)$, the \emph{query evaluation problem} is to decide whether $\mu \in \semm {\mathbf{q}(?x, ?y)}{G}$, that is, whether the tuple $\mu$ is in the result of the query $\mathbf{q}$ on the RDF graph $G$.
There are two kinds of computational complexity in the query evaluation problem \cite{abs_book,ahv_book}:
\begin{compactitem}
\item the \emph{data complexity} refers to the complexity w.r.t. the size of the RDF graph $G$, given a fixed query $\mathbf{q}$; and
\item the \emph{combined complexity} refers to the complexity w.r.t. the size of query $\mathbf{q}$ and the RDF graph $G$.
\end{compactitem}

A CFG $\G = (N, A, R)$ is said to be in \emph{norm form} if all of its production rules are of the form $v \to uw$, $v \to a$, or $v \to \epsilon$ where $v, u, w \in N$ and $a \in A$. Note that this norm form deviates from the usual \emph{Chomsky Normal Form} \cite{chomsky_nf} where the start non-terminals are absent. Indeed, every CFG is equivalent to a CFG in norm form, that is, for every CFG $\G$, there exists some CFG $\G'$ in norm form constructed from $\G$ in polynominal time such that $\LL(\G_{v}) = \LL(\G'_{v})$ for every $v \in N$ \cite{DBLP:conf/icdt/Hellings14}.

Let $G$ be an RDF graph and $\G = (N, A, R)$ a CFG. Given a non-terminal $v \in N$, let $\mathcal{R}_v(G)$ be \emph{the context-free relation} of $G$ w.r.t. $v$ can be defined as follows:
\begin{equation}
\mathcal{R}_v(G):= \{(a, b)\mid \exists\, \pi = (a c_1 \ldots c_m b) \text{ a path in } G, \Trace(\pi) \cap \LL(\G_{v}) \neq \emptyset\}.
\end{equation}

Conveniently, the query evaluation of CCFPQ over an RDF graph can be reduced into the conjunctive first-order query over the context-free relations. Based on the conjunctive context-free recognizer for graphs presented in \cite{DBLP:conf/icdt/Hellings14}, we  directly obtain a conjunctive context-free recognizer (see Algorithm \ref{alg:recognizer}) for RDF graphs by adding a convertor to transform an RDF graph into an edge-labeled directed graph (see Algorithm \ref{alg:convertor}).

\vspace*{-0.5cm}
\begin{algorithm}[h]
\centering
\begin{algorithmic}[1]
 \REQUIRE $G$: an RDF graph; $\G = (N, A, R)$: a CFG in norm form; $v \in N$.
 \ENSURE $\{(v, a, b) \mid (a, b) \in \mathcal{R}_v(G)\}$
\STATE $\Theta:= \{(v, a, a) \mid (a \in \voc(G)) \wedge (v \to \epsilon \in P)\}$
\STATE $\quad\quad\cup \{(v, a, b) \mid ((a, l, b) \in \vocvertor(G)) \wedge (v \to l) \in P\}$
  \STATE $\Theta_{new}: = \Theta$
 \WHILE{$\Theta_{new} \neq \emptyset$}
  \STATE pick and remove a $(v, a, b)$ from $\Theta_{new}$
  \FORALL{$(u, a', a) \in \Theta$}
    \FORALL{$v' \to uv \in R \wedge ((v', a', b) \not\in \Theta)$}
     \STATE  $\Theta_{new}:= \Theta_{new} \cup \{(v', a', b)\}$
     \STATE  $\Theta := \Theta \cup \{(v', a', b)\}$
    \ENDFOR
 \ENDFOR
  \FORALL{$(u, b, b') \in \Theta$}
    \FORALL{$u' \to vu \in R \wedge ((u', a, b') \not\in \Theta)$}
    \STATE $\Theta_{new}:= \Theta_{new} \cup \{(u', a, b')\}$
    \STATE $\Theta := \Theta \cup \{(u', a, b')\}$
 \ENDFOR
 \ENDFOR
 \ENDWHILE
 \RETURN $\Theta$
 \caption{Conjunctive context-free recognizer for RDF}\label{alg:recognizer}
 \end{algorithmic}
\end{algorithm}

\begin{algorithm}[h]
\centering
\begin{algorithmic}[1]
 \REQUIRE {$G$: an RDF graph}
 \ENSURE {$\vocvertor(G) = (\mathcal{V}, \mathcal{E})$}
  \STATE {$\mathcal{V}:= \voc(G)$}
 \STATE {$\mathcal{E}:= \{(c, \self, c), (c, \self::c, c) \mid c \in \voc(G)\}$}
 \STATE {$G_{new}: = G$}
 \WHILE{$G_{new} \neq \emptyset$}
 \STATE pick and remove a triple $(s, p, o)$ from $G_{new}$
 \STATE {$\mathcal{E}:= \mathcal{E} \cup \{(s, \nexts::p, o), (s, \nexts, o)$,
$(o, \nexts^{-1}::p, s), (o, \nexts^{-1}, s)$, \\
$\quad \quad \quad \quad (s, \edges::o, p), (s, \edges, p)$,
$(p, \edges^{-1}::o, s), (p, \edges^{-1}, s)$,\\
$\quad \quad \quad \quad (p, \nodes::s, o), (p, \nodes, o)$,
$(o,\nodes^{-1}::s, p), (o, \nodes^{-1}, p)\}$
 }
 \ENDWHILE
 \RETURN $\vocvertor(G)$
 \caption{RDF convertor}\label{alg:convertor}
\end{algorithmic}
\end{algorithm}

Given a path $\pi$ and a context-free grammar $\G$, Algorithm \ref{alg:recognizer} is sound and complete to decide whether the path $\pi$ in RDF graphs has a trace generated from the grammar $\G$.
\begin{proposition}\label{prop:ccfpq-RDF}
Let $G$ be an RDF graph and $\G = (N, A, R)$ a CFG in norm form. For every $v \in N$, let $\Theta$ be the result computed in Algorithm \ref{alg:recognizer}, $(v, a, b) \in \Theta$ if and only if $(a, b) \in \mathcal{R}_v(G)$.
\end{proposition}
%
%

Moreover, we can easily observe the worst-case complexity of Algorithm \ref{alg:recognizer} since the complexity of Algorithm \ref{alg:convertor} is $\mathcal{O}(|G|)$.
\begin{proposition}\label{prop:recognizer_complexity}
Let $G$ be an RDF graph and $\G = (N, A, R)$ a CFG. Algorithm \ref{alg:recognizer} applied to $G$ and $\G$ has a worst-case complexity of $\mathcal{O}((|N||G|)^3)$.
\end{proposition}

As a result, we can conclude the following proposition.
\begin{proposition}\label{prop:ccfpq_complexity}
The followings hold:
\begin{enumerate}
\item The query evaluation of CCFPQ has polynomial data complexity;
\item The query evaluation of CCFPQ has NP-complete combined complexity.
\end{enumerate}
\end{proposition}
%

\paragraph{\bf Union of CCFPQ}

An extension of CCFPQ capturing more expressive power such as disjunctive
capability is introducing the union of CCFPQ. For instance, given a gene (e.g., \emph{Gene(B)}) in the biomedical ontology (see Figure \ref{fig:toy}), we wonder to find those genes which are relevant to this gene, that is, those genes either are similar to it (e.g., \emph{Gene(C)}) or belong to the same pathway (e.g., \emph{Gene(S)}).

A \emph{union of conjunctive context-free path query} (UCCFPQ) is of the form
\begin{equation}\label{equ:uccftp}
\mathbf{q}(?x, ?y):= \bigvee^{m}_{i=1}\, \mathbf{q}_i(?x, ?y),
\end{equation}
where $\mathbf{q}_i(?x, ?y)$ is a CCFPQ for all $i = 1, \ldots, m$.

Analogously, we can define \emph{union of simple conjunctive context-free path query} written by $UCCFPQ^{s}$.

Semantically, let $G$ be an RDF graph, we define 
\begin{equation}\label{equ:uccftp-semantics}
\semm {\mathbf{q}(?x, ?y)}{G} = \bigcup^{m}_{i=1}\, \semm {\mathbf{q}_i(?x, ?y)}{G},
\end{equation}
where $\semm {\mathbf{q}_i(?x, ?y)}{G}$ is defined as the semantics of CCFPQ for all $i = 1, \ldots, m$.

In Example \ref{exam:bio}, based on $\G = (N, A, R)$, we construct a CFG $\G' = (N', A', R')$ where $N' = N \cup \{s\}$, $A = A\cup \{\nexts::\emph{belongsTo}$, $ \nexts^{-1}::\emph{belongsTo}\}$, and $R' = R\cup \{s \to (\nexts::\emph{belongsTo})s(\nexts^{-1}::\emph{belongsTo})\}$.
Consider a UCCFPQ $\mathbf{q}(?x, ?y) := v(?x, ?y)\, \vee\, s(?x,?y)$,
$\semm {\mathbf{q}(?x, ?y)}{G_{\mathrm{bio}}} = \{(?x =\emph{Gene(B)}, ?y = \emph{Gene(C)})$, $(?x =\emph{Gene(B)}, ?y =\emph{Gene(S)})\}$. That is, $\semm {\mathbf{q}(?x, ?y)}{G_{\mathrm{bio}}}$ returns all pairs where the first gene is relevant to the latter.

Note that the query evaluation of UCCFPQ has the same complexity as
that of the evaluating of CCFPQ since we can simply evaluate a number (linear in the
size of a UCCFPQ) of CCFPQs in isolation \cite{ahv_book}.

\section{Expressivity of (U)(C)CFPQ}\label{sec:expressivity}
In this section, we investigate the expressivity of (U)(C)CFPQ by referring to nested regular expressions \cite{nsparql} and fragments of nre.

We discuss the relations between variants of UCCFPQ and variants of (nested) regular expressions and obtain the following results:
\begin{compactenum}
\item nre$_{0}$-triples can be expressed in CFPQ;
\item nre$_{0}(\mathrm{N})$-triples can be expressed in CCFPQ;
\item nre$_{0}(|)$-triples can be expressed in UCCFPQ$^{s}$;
\item nre-triples can be expressed in UCCFPQ.
\end{compactenum}

\paragraph{\bf 1. nre$_{0}$ in CFPQ}


The following proposition shows that CFPQ can express nre$_{0}$-triples.
\begin{proposition}\label{prop:nre0-CFPQ}
For every nre$_{0}$-triple $(?x, e, ?y)$, there exist some CFG $\G =(N, A, R)$ and some CFPQ $\mathbf{q}(?x, ?y)$  such that for every RDF graph $G$, we have $\semm {(?x, e, ?y)}{G} =\semm {\mathbf{q}(?x, ?y)}{G}$.
\end{proposition}

\paragraph{\bf 2. nre$_{0}(\mathrm{N})$ in CCFPQ}


Let $\G$ be a CFG. A \emph{CCFPQ} $\mathbf{q}(?x, ?y)$ is in \emph{nested norm form} if the following holds:
\begin{equation}\label{equ:nnf}
\mathbf{q}(?x, ?y):= ((?x', ?y', ?z') \wedge v(?x, ?y)) \wedge \mathbf{q}_1(?u, ?w),
\end{equation}
where
\begin{compactitem}
\item $\{?x, ?y\} \cap \{?x', ?y', ?z'\} \neq \emptyset$;
\item $\{?x', ?y', ?z'\} \cap \{?u, ?w\} \neq \emptyset$;
\item $\mathbf{q}_1(?u, ?w)$ is a CCFPQ.
\end{compactitem}
Note that $(?x', ?y', ?z')$ is used to express a nested nre of the form $\axis::[e]$ and $v(?x, ?y)$ is necessary to express a nested nre of the form $\self::[e]$.

The following proposition shows that CCFPQ can express nre$_{0}(\mathrm{N}$)-triples.
\begin{proposition}\label{prop:nre-nested-CCFPQ}
For every nre$_{0}(\mathrm{N}$)-triple $(?x, e, ?y)$, there exist a CFG $\G =(N, A, R)$ and a CCFPQ $\mathbf{q}(?x, ?y)$ in nested norm form (\ref{equ:nnf}) such that for every RDF graph $G$, we have $\semm {(?x, e, ?y)}{G} =\semm {\mathbf{q}(?x, ?y)}{G}$.
\end{proposition}

\paragraph{\bf 3. nre$_{0}(|)$ in UCCFPQ$^{s}$}

Let $e$ be an nre. We say $e$ is in \emph{union norm form} if $e$ is of the following form $e_1 | e_2 | \ldots | e_m$
where $e_i$ is an nre$_{0}(\mathrm{N}$) for all $i=1,\ldots, m$.

We can conclude that each nre-triple is equivalent to an nre in union norm form.
\begin{proposition}\label{prop:nre-norm form}
For every nre-triple $(?x, e, ?y)$, there exists some $e'$ in union norm form such that $\semm {(?x, e, ?y)}{G} = \semm {(?x, e', ?y)}{G}$ for every RDF graph $G$.
\end{proposition}

The following proposition shows that UCCFPQ$^{s}$ can express nre$_{0}(|)$.
\begin{proposition}\label{prop:nre-union-UCCFPQs}
For every nre$_{0}(|)$-triple $(?x, e, ?y)$, there exists some CFG $\G =(N, A, R)$ and some UCCFPQ$^{s}$ $\mathbf{q}(?x, ?y)$ in nested norm form such that for every RDF graph $G$, we have $\semm {(?x, e, ?y)}{G} =\semm {\mathbf{q}(?x,?y)}{G}$.
\end{proposition}
%
%

\paragraph{\bf 4. nre in UCCFPQ}

By Proposition \ref{prop:nre-nested-CCFPQ} and Proposition \ref{prop:nre-union-UCCFPQs}, we can conclude that
\begin{proposition}\label{prop:nre-UCCFPQ}
For every nre-triple $(?x, e, ?y)$, there exists some CFG $\G =(N, A, R)$ and some UCCFPQ $\mathbf{q}(?x, ?y)$ in nested norm form such that for every RDF graph $G$, we have $\semm {(?x, e, ?y)}{G} =\semm {\mathbf{q}(?x,?y)}{G}$.
\end{proposition}

However, those results above in this subsection are not vice versa since the context-free language is not expressible in any nre.
\begin{proposition}\label{prop:CFPQ-nre}
CFPQ is not expressible in any nre.
\end{proposition}

\section{Context-free SPARQL}\label{sec:cfSPARQL}
In this section, we introduce an extension language \emph{context-free SPARQL} (for short, \emph{cfSPARQL}) of SPARQL by using context-free triple patterns, plus SPARQL basic operators $\UNION$, $\andd$, $\OPT$, $\FILTER$, and $\SELECT$ and its expressiveness.

A \emph{context-free triple pattern} (cftp) is of the form $(?x, \mathbf{q}, ?y)$ where $\mathbf{q}(?x, ?y)$ is a CFPQ. Analogously, we can define \emph{union of conjunctive context-free triple pattern} (for short, \emph{uccftp}) by using UCCFPQ.

\paragraph{\textbf{cfSPARQL and query evaluation}} Formally, cfSPARQL (graph) patterns are
then recursively constructed from context-free triple patterns:
\begin{compactitem}
\item A cftp is a cfSPARQL pattern;
\item A triple pattern of the form $(?x, ?y, ?z)$ is a cfSPARQL pattern;
\item All $P_1\, \UNION\, P_2$, $P_1\, \andd\, P_2$, and $P_1\, \OPT\, P_2$ are cfSPARQL patterns if $P_1, P_2$ are cfSPARQL patterns;
\item $P\, \FILTER\, C$ if $P$ is a cfSPARQL pattern and $C$ is a contraint;
\item $\SELECT_{S}(P)$ if $P$ is a cfSPARQL pattern and $S$ is a set of variables.
\end{compactitem}

\begin{remark}\label{remark:xyz}
In cfSPARQL, we allow triple patterns of form $(?x, ?y, ?z)$ (see Item 2), which can express any SPARQL triple pattern together with $\FILTER$ \cite{DBLP:journals/corr/ZhangB14}, to ensure that SPARQL is still expressible in cfSPARQL while SPARQL is not expressible in nSPARQL since any triple pattern $(?x, ?y, ?z)$ is not expressible in nSPARQL \cite{nsparql}. Our generalization of nSPARQL inherits the power of queries without more cost and maintains the coherence between CFPQ and ``nested'' nre of the form $\axis::[e]$. Moreover, this extension in cfSPARQL coincides with our proposed CCFPQ where triple patterns of the form $(?x, ?y, ?z)$ are allowed.
\end{remark}

Semantically, let $P$ be a cfSPARQL pattern and $G$ an RDF graph, $\semm {(?x, \mathbf{q}, ?y)}{G}$ is defined as $\semm {\mathbf{q}(?x, ?y)}{G}$ and other expressive cfSPARQL patterns are defined as normal \cite{nsparql,perez_sparql_tods}.

\begin{proposition}\label{prop:SPARQL-cfSPARQL}
SPARQL is expressible in cfSPARQL but not vice versa.
\end{proposition}

A \emph{cfSPARQL query} is a pattern. 

We can define \emph{union of conjunctive context-free SPARQL query} (for short, \emph{uccfSPARQL}) by using uccftp in the analogous way.

At the end of this subsection, we discuss the complexity of evaluation problem of uccfSPARQL queries.

For a given RDF graph $G$, a uccftp $P$, and a mapping $\mu$, the query evaluation problem is to decide whether $\mu$ is in $\semm {P}{G}$.

\begin{proposition}\label{prop:uccfSPARQL-complexity}
The evaluation problem of uccfSPARQL queries has the same complexity as the evaluation problem of SPARQL queries.
\end{proposition}

As a direct result of Proposition \ref{prop:nre-UCCFPQ}, we can conclude
\begin{corollary}
nSPARQL is expressible in uccfSPARQL but not vice versa.
\end{corollary}

\paragraph{\textbf{On the expressiveness of cfSPARQL}} In this subsection, we show that cfSPARQL has the same expressiveness as uccfSPARQL. In other words, cfSPARQL
is enough to express UCCFPQ on RDF graphs.

Since every cfSPARQL pattern is a uccfSPARQL pattern, we merely show that uccfSPARQL is expressible in cfSPARQL.
\begin{proposition}\label{prop:uccfSPARQL-cfSPARQL}
For every uccfSPARQL pattern $P$, there exists some cfSPARQL pattern $Q$ such that $\semm {P}{G} = \semm {Q}{G}$ for any RDF graph $G$.
\end{proposition}
%
%

%

\section{Relations on (nested) regular expressions with negation}\label{sec:nre_neg}
In this section, we discuss both the relation between UCCFPQ and nested regular expressions with negation and the relation between cfSPARQL and variants of nSPARQL.

\paragraph{\textbf{Nested regular expressions with negation}} \emph{A nested regular
expression with negation (nre$^{\neg}$)} is an extension of nre by adding
two new operators ``difference ($e_1 - e_2$)'' and ``negation ($e^{c}$)''
\cite{Taski_SPARQL}.


Semantically, let $e, e_1, e_2$ be three nre$^{\neg}$s and $G$ an RDF graph,
\begin{compactitem}
\item $\sem {e_{1} - e_{2}}  = \{(a, b) \in \sem {e_{1}} \mid (a, b) \not \in \sem {e_{2}}\}$;
\item $\sem {e^{c}}  =  \{(a, b) \in \voc(G) \times \voc(G) \mid  (a, b) \not\in \semm {e}{G}\}$.
\end{compactitem}

Analogously, an nre$^{\neg}$-triple pattern is of the form $(?x, e, ?y)$ where $e$ is an nre$^{\neg}$. Clearly, nre$^{\neg}$-triple pattern is non-monotone.

Since nre is monotone, nre is strictly subsumed in nre$^{\neg}$ \cite{Taski_SPARQL}.
Though property paths in SPARQL 1.1 \cite{sparql1.1,DBLP:journals/jancl/PolleresW13} are not expressible in nre since property paths allow the negation of IRIs, property paths can be still expressible in the following subfragment of nre$^{\neg}$: let $c, c_1, \ldots, c_{n+m} \in U$,
\begin{multline*}\label{equ:pp}
e := \nexts::c \mid e / e \mid \self::[e] \mid e^{\ast} \mid e^{+} \mid \nexts^{-1}::[e] \mid\\ (\nexts::c_1 | \ldots | \nexts::c_n | \nexts^{-1}::c_{n+1} | \ldots | \nexts^{-1}::c_{n+m})^c.
\end{multline*}
Note that $e^{+}$ can be expressible as the expression $e^{\ast} - \self$.
%
\begin{proposition}\label{prop:uccftp-nre_negation}
uccftp is not expressible in any nre$^{\neg}$-triple pattern.
\end{proposition}

Due to the non-monotonicity of nre$^{\neg}$, we have that nre$^{\neg}$ is beyond the expressiveness of any union of conjunctive context-free triple patterns even the star-free nre$^{\neg}$ (for short, sf-nre$^{\neg}$) where the Kleene star ($\ast$) is not allowed in nre$^{\neg}$.
\begin{proposition}\label{prop:nre_negation-uccftp}
sf-nre$^{\neg}$-triple pattern is not expressible in any uccftp.
\end{proposition}
%

In short, nre$^{\neg}$-triple pattern and uccftp cannot express each other. Indeed, negation could make the evaluation problem hard even allowing a limited form of negation such as property paths \cite{DBLP:journals/tods/LosemannM13}.

\paragraph{\textbf{cfSPARQL can express nSPARQL$^{\neg}$}} Following nSPARQL, we can
analogously construct the language nSPARQL$^{\neg}$ which is built on
nre$^{\neg}$, by adding SPARQL operators $\UNION$, $\andd$, $\OPT$, $\FILTER$, and
$\SELECT$.

Though uccftps cannot express nre$^{\neg}$-triple patterns by Proposition \ref{prop:uccftp-nre_negation}, cfSPARQL can express nSPARQL$^{\neg}$ since nSPARQL$^{\neg}$ is still expressible in nSPARQL \cite{Taski_SPARQL}.

\begin{corollary}\label{prop:cfSPARQL-nSPARQL}
nSPARQL$^{\neg}$ is expressible in cfSPARQL.
\end{corollary}

\subsection{Overview}
Finally, Figure \ref{fig:overviewofCFPQ} and Figure \ref{fig:overviewofcfSPARQL} provide the implication of the results on RDF graphs for the general relations between variants of CFPQ and nre and the general relations between cfSPARQL and nSPARQL where $\mathcal{L}_1 \to \mathcal{L}_2$ denotes that $\mathcal{L}_1$ is expressible in $\mathcal{L}_2$ and $\mathcal{L}_1 \leftrightarrow \mathcal{L}_2$ denotes that $\mathcal{L}_1 \to \mathcal{L}_2$ and $\mathcal{L}_2 \to \mathcal{L}_1$. Analogously, nSPARQL$^{\mathrm{sf}}$ is an extension of SPARQL by allowing star-free nre$^\neg$-triple patterns.
\begin{figure}[h]
\setlength{\abovecaptionskip}{0.25cm}
\setlength{\belowcaptionskip}{-0.65cm}
\centering
\begin{minipage}[t]{0.5\linewidth}
\begin{tikzpicture}[>=latex]
  %
  %
  \tikzstyle{state} = [minimum height=1.5em, minimum width=1.5em, node distance=4em, font={\sffamily}]
  \tikzstyle{stateEdgePortion} = [black,thick];
  \tikzstyle{stateEdge} = [stateEdgePortion, ->];
  \tikzstyle{stateEdgeD} = [stateEdgePortion,dashed,->];
  \tikzstyle{stateEdgeUD} = [stateEdgePortion,dashed];
  \tikzstyle{edgeLabel} = [pos=0.5, text centered, font={\sffamily\small}];

  %
  %
  \node[state, name=p1] {UCCFPQ};
  \node[state, name=p2, right of=p1, xshift = 1.1em] {nre$^{\neg}$};
  \node[state, name=p3, below of=p1, xshift = -5em] {CCFPQ};
  \node[state, name=p4, below of=p1] {nre};
  \node[state, name=p5, below of=p1, xshift = 5em] {UCCFPQ$^s$};
  \node[state, name=p6, below of=p3] {nre$_{0}(\mathrm{N})$};
  \node[state, name=p7, below of=p4] {CFPQ};
  \node[state, name=p8, below of=p5] {nre$_{0}(|)$};
  \node[state, name=p9, below of=p7] {nre$_{0}$};

  \draw ($(p4.north)$) edge[stateEdge] ($(p2.south)$);
  \draw ($(p3.north)$) edge[stateEdge] ($(p1.south)$);
  \draw ($(p4.north)$) edge[stateEdge] ($(p1.south)$);
  \draw ($(p5.north)$) edge[stateEdge] ($(p1.south)$);
  \draw ($(p6.north)$) edge[stateEdge] ($(p3.south)$);
  \draw ($(p6.north)$) edge[stateEdge] ($(p4.south)$);
  \draw ($(p7.north)$) edge[stateEdge] ($(p3.south)$);
  \draw ($(p7.north)$) edge[stateEdge] ($(p5.south)$);
  \draw ($(p8.north)$) edge[stateEdge] ($(p4.south)$);
  \draw ($(p8.north)$) edge[stateEdge] ($(p5.south)$);
  \draw ($(p9.north)$) edge[stateEdge] ($(p6.south)$);
  \draw ($(p9.north)$) edge[stateEdge] ($(p7.south)$);
  \draw ($(p9.north)$) edge[stateEdge] ($(p8.south)$);
\end{tikzpicture}
\caption{Known relations between variants\\ of CFPQ and variants of  nre.}\label{fig:overviewofCFPQ}
\end{minipage}%
\begin{minipage}[t]{0.5\linewidth}
\begin{tikzpicture}[>=latex]

  %
  %
  \tikzstyle{state} = [minimum height=2em, minimum width=2em, node distance=4em, font={\sffamily}]
  \tikzstyle{stateEdgePortion} = [black,thick];
  \tikzstyle{stateEdge} = [stateEdgePortion, ->];
  \tikzstyle{stateEdgeD} = [stateEdgePortion,stealth-stealth];
  \tikzstyle{edgeLabel} = [pos=0.5, text centered, font={\sffamily\small}];

  %
  %
  \node[state, name=p1] {cfSPARQL};
  \node[state, name=p2, right of=p1, xshift=6em] {uccfSPARQL};
  \node[state, name=p4, below of=p2] {nSPARQL$^{\neg}$};
  \node[state, name=p5, below of=p4] {nSPARQL};
  \node[state, name=p6, below of=p1] {SPARQL};
  \node[state, name=p7, below of=p6] {nSPARQL$^{\mathrm{sf}}$};

  \draw ($(p6.north)$) edge[stateEdge] ($(p1.south)$);
  \draw ($(p4.north)$) edge[stateEdge] ($(p2.south)$);
  \draw ($(p1.east)$) edge[stateEdgeD] ($(p2.west)$);
  \draw ($(p5.north)$) edge[stateEdge] ($(p4.south)$);
  \draw ($(p7.north)$) edge[stateEdge] ($(p6.south)$);
  \draw ($(p7.east)$) edge[stateEdge] ($(p5.west)$);
\end{tikzpicture}
\caption{Known relations between variants\\ of cfSPARQL and variants of  nSPARQL.}\label{fig:overviewofcfSPARQL}
 \end{minipage}%
\end{figure}

\vspace*{-0.8cm}

\section{Implementation and evaluation}\label{sec:implement}
\vspace*{-0.2cm}
In this section, we have implemented the two algorithms for CFPQs without any optimization. Two context-free path queries over RDF graphs were evaluated and we found some results which cannot be captured by any regular expression-based path queries from RDF graphs.

The experiments were performed under Windows 7 on a Intel i5-760, 2.80GHz
CPU system with 6GB memory. The program was written in Java 7 with maximum
2GB heap space allocated for JVM. Ten popular ontologies like \textit{foaf}, \textit{wine}, and \textit{pizza} were used for testing.

\vspace*{-0.3cm}
\paragraph{\textbf{Query 1}}
Consider a CFG $\G_1 = (N, A, R)$ where $N = \{S\}$, $A =
\{\nexts^{-1}::\textit{subClassOf},
\nexts::\textit{subClassOf},\nexts^{-1}::\textit{type},
\nexts::\textit{type}\}$, and $R = \{S \to
(\nexts^{-1}::\textit{subClassOf})\,S\,(\nexts::\textit{subClassOf}),S \to
(\nexts^{-1}::\textit{type})\,S\,(\nexts::\textit{type}),S \to
\varepsilon\}$. The query $Q_1$ based on the grammar $\G_1$ can return all
pairs of concepts or individuals at the same layer of the hierarchy of RDF
graphs. Table \ref{fig:res_queries} shows the experimental results of $Q_1$
over the testing ontologies. Note that $\#results$ denotes that number of
pairs of concepts or individuals corresponding to $Q_1$.

Taking the ontology \textit{foaf}, for example, the query $Q_1$ over
\textit{foaf} returns pairs of concepts like $(\textit{foaf:Document},
\textit{foaf:Person})$, which shows that the two concepts,
\textit{Document} and \textit{Person}, are at the same layer of the
hierarchy of \textit{foaf}, where the top concept ($\textit{owl:Thing}$) is
at the first layer.

\paragraph{\textbf{Query 2}}
Similarly, consider a CFG $\G_2 = (N, A, R)$ where $N =
\{S,B\}$, $A =
\{\nexts^{-1}::\textit{subClassOf},\nexts::\textit{subClassOf}\}$, and $R =
\{S \to BS,B \to
(\nexts::\textit{subClassOf})\,B\,(\nexts^{-1}::\textit{subClassOf}),B \to
B(\nexts^{-1}::\emph{subClassOf})\,B \to
(\nexts::\textit{subClassOf})(\nexts^{-1}::\textit{subClassOf}),S \to
\varepsilon\}$. The query $Q_2$ based on the grammar $\G_2$ can return all
pairs of concepts which are at adjacent two layers of the hierarchy of RDF
graphs. We also take the ontology \textit{foaf}, for example, the query
$Q_2$ over \textit{foaf} returns pairs of concepts like
($\textit{foaf:Person}, \textit{foaf:PersonalProfileDocument}$), which
denotes that \textit{Person} is at higher layer than
\textit{PersonalProfileDocument}, since \textit{PersonalProfileDocument} is
a subclass of \textit{Document}. Table \ref{fig:res_queries} shows the
experimental results of $Q_2$ over the testing ontologies.

\vspace*{-0.8cm}
%
\begin{table}[H]
	\centering
	{	
		\caption{{The evaluation results of $Q_1$ and $Q_2$}}	
		\scalebox{1.0}{
			\centering
			\renewcommand{\arraystretch}{1.5}
			\begin{tabular}{|c|c|c|c|c|c| }
				\hline
				\textbf{Ontology} & \textbf{\#triples} & \multicolumn{2}{c|}{\textbf{Query 1}} & \multicolumn{2}{c|}{\textbf{Query 2}}\\ \cline{3-6}
                & &\textbf{time(ms)} & \textbf{\#results}¡¡& \textbf{time(ms)} & \textbf{\#results} \\
				\hhline{|=|=|=|=|=|=|}
				protege&41&468&509&5&0 \\
				\hline
				funding&144&499&296&125&77\\
				\hline
				skos&254&1044&810&16&1\\
				\hline
				foaf&454&5027&1929&1154&324\\
				\hline
				generation&319&6091&2164&13&0\\
				\hline
				univ-bench&306&20981&2540&532&228\\
				\hline
				travel&327&13971&2499&281&151\\
				\hline
				people+pets&703&82081&9472&247&120\\
				\hline
				biomedical-measure-primitive&459&420604&15156&1068851&9178\\
				\hline
				atom-primitive&561&515285&15454&4711499&13940\\
				\hline
				pizza&1980&3233587&56195&255853&4694\\
				\hline
				wine&2012&4075319&66572&273&79\\
				\hline
			\end{tabular}
		}
		
		\label{fig:res_queries}
	}
\end{table}

\vspace*{-0.8cm}

\section{Conclusions}\label{sec:conclusion}
In this paper, we have proposed context-free path queries (including some
variants) to navigate through an RDF graph and the context-free SPARQL query
language for RDF built on context-free path queries by adding the standard
SPARQL operators. Some investigation about some fundamental properties of those
context-free path queries and their context-free SPARQL query languages has been presented. We proved that CFPQ, CCFPQ, UCCFPQ$^{s}$, and UCCFPQ strictly express basic
nested regular expression (nre$_{0}$), nre$_{0}(\mathrm{N})$,
nre$_{0}(|)$, and nre, respectively. Moreover, uccfSPARQL has the same
expressiveness as cfSPARQL; and both SPARQL and nSPARQL are expressible in
cfSPARQL. Furthermore, we looked at the relationship between context-free path
queries and nested regular expressions with negation (which can express
property paths in SPARQL1.1) and the relationship between cfSPARQL queries and
nSPARQL queries with negation (nSPARQL$^{\neg}$). We found that neither CFPQ
nor nre$^{\neg}$ can express each other while nSPARQL$^{\neg}$ is still
expressible in cfSPARQL. Finally, we discussed the query evaluation problem
of CFPQ and cfSPARQL on RDF graphs. The query evaluation of UCCFPQ maintains
the polynomial time data complexity and NP-complete combined complexity the same as conjunctive first-order queries and the query evaluation of
cfSPARQL maintains the complexity as the same as SPARQL. These results
provide a starting point for further research on expressiveness of
navigational languages for RDF graphs and the relationships among regular path
queries, nested regular path queries, and context-free path queries on RDF
graphs.

There are a number of practical open problems. In this paper, we restrict
that RDF data does not contain \emph{blank nodes} as the same treatment in
nSPARQL. We have to admit that blank nodes do make RDF data more expressive
since a blank node in RDF is taken as an existentially quantified variable
\cite{DBLP:journals/ws/HoganAMP14}.  An interesting future work is to extend
our proposed (U)(C)CFPQ for general RDF data with blank nodes by allowing
path variables which are already valid in some extensions of SPARQL such as
SPARQ2L\cite{DBLP:conf/www/AnyanwuMS07},
SPARQLeR\cite{DBLP:conf/esws/KochutJ07}, PSPARQL
\cite{DBLP:journals/ws/AlkhateebBE09}, and CPSPARQL
\cite{DBLP:conf/swws/AlkhateebBE08,DBLP:journals/ijwis/AlkhateebE14}, which
are popular in querying general RDF data with blank nodes.

\vspace*{-0.4cm}
\section*{Acknowledgments}
\vspace*{-0.2cm}
The authors thank Jelle Hellings and Guohui Xiao for their helpful and
constructive comments. This work is supported by the program of the National Key Research and Development Program of China (2016YFB1000603) and the National Natural Science Foundation of China (NSFC) (61502336, 61373035). Xiaowang Zhang is supported by Tianjin Thousand Young Talents Program.

%
\end{document}